\documentclass[sigconf]{acmart}
\usepackage{svg}

\usepackage[utf8]{inputenc}
\usepackage{textgreek}
\usepackage{booktabs} 
\usepackage{graphicx} 
\usepackage{csquotes}
\usepackage{array} 
\usepackage[none]{hyphenat}
\let\ket\relax
\usepackage[braket, qm]{qcircuit}
\usepackage{physics}
\usepackage{hyperref}
\usepackage{fancyhdr}
\AtBeginDocument{%
  }

\setcopyright{acmlicensed}
\copyrightyear{2025}
\acmYear{2025}
\setcopyright{cc}
\setcctype{by}
\acmConference[QSec '25]{Proceedings of the 2025 Quantum Security and Privacy Workshop}{October 13--17, 2025}{Taipei, Taiwan}
\acmBooktitle{Proceedings of the 2025 Quantum Security and Privacy Workshop (QSec '25), October 13--17, 2025, Taipei, Taiwan}\acmDOI{10.1145/3733825.3765282}
\acmISBN{979-8-4007-1913-4/2025/10}

\begin{document}
\sloppy

\title{Pulse-to-Circuit Characterization of Stealthy Crosstalk Attack on Multi-Tenant Superconducting Quantum Hardware}
\author{Syed Emad Uddin Shubha, Tasnuva Farheen\\Division of Computer Science, Louisiana State University}

\begin{abstract}
Hardware crosstalk in multi-tenant superconducting quantum computers constitutes a significant security threat, enabling adversaries to inject targeted errors across tenant boundaries. We present the first end-to-end framework for mapping physical pulse-level attacks to interpretable logical error channels, integrating density-matrix simulation, quantum process tomography (QPT), and a novel isometry-based circuit extraction method. Our pipeline reconstructs the complete induced error channel and fits an effective logical circuit model, revealing a fundamentally asymmetric attack mechanism: one adversarial qubit acts as a driver to set the induced logical rotation, while a second, the catalyst, refines the attack’s coherence. Demonstrated on a linear three-qubit system, our approach shows that such attacks can significantly disrupt diverse quantum protocols, sometimes reducing accuracy to random guessing, while remaining effective and stealthy even under realistic hardware parameter variations. We further propose a protocol-level detection strategy based on observable attack signatures, showing that stealthy attacks can be exposed through targeted monitoring and providing a foundation for future defense-in-depth in quantum cloud platforms.
\end{abstract}



\maketitle
\section{Introduction}

Quantum computing has rapidly advanced, enabling transformative capabilities in cryptography, optimization, and quantum simulation. As this technology scales, particularly on cloud-based, multi-tenant platforms, new security vulnerabilities are emerging at the physical layer. The power of quantum computation relies on the ability to couple qubits to perform entangling gates; however, this same engineered coupling creates an inherent security risk: \emph{hardware crosstalk}. 
It arises when control pulses or interactions intended for one qubit unintentionally influence its neighbors. This interference, often manifesting as persistent, always-on interactions like ZZ, ZX, or YX couplings~\cite{mundada2019suppression}, is not merely a barrier to computational fidelity. As we will demonstrate, it is a vector for sophisticated security exploits that has only recently begun receiving systematic attention.

Historically, research into quantum hardware security has approached crosstalk from two primary directions. Traditionally treated as a calibration or fidelity problem~\cite{ash2020analysis, zhao2022quantum, li2022pulse, rudinger2021experimental, sarovar2020detecting}, the first perspective views crosstalk purely as noise requiring mitigation. Foundational studies have applied fault injection, gate-set tomography, and randomized benchmarking to characterize its detrimental effects. While hardware engineers focus on this fidelity challenge, recent work has begun to integrate Quantum Process Tomography (QPT) to analyze a broader spectrum of coupling mechanisms, though these efforts often do not fully characterize the underlying physical-to-logical mapping or consider asymmetric qubit roles~\cite{zhou2025characterization}.

The second research direction explicitly investigates crosstalk as an \emph{active attack mechanism}~\cite{ash2020analysis, maurya2024understanding, choudhury2024crosstalk}. By injecting tailored pulses into their assigned qubits, an attacker can induce targeted errors in a victim’s computation, such as flipping the phase of a target qubit to alter a superposition state~\cite{xu2024jailbreaking}. Critically, such attacks can degrade computational integrity, bias outcomes, or introduce errors without requiring direct access to the victim’s circuit, making them a serious threat to the security of shared quantum resources. These active attack studies can be broadly categorized into two approaches. The first involves an adversary who directly modifies the pulse-level definitions within a victim's circuit, exploiting software and interface vulnerabilities~\cite{xu2024jailbreaking}. The second, more subtle approach uses crosstalk as a passive side-channel for \emph{confidentiality attacks}, where adversaries observe correlated error rates on their own qubits to infer properties of a victim's computation~\cite{choudhury2024crosstalk, maurya2024understanding}.

Despite these advances, a critical research gap persists. Prior work has largely overlooked the potential of crosstalk as an active vector for \emph{integrity attacks}. In this work, we explore this different threat: using crosstalk as an \textit{active injection vector}, where an adversary does not tamper with the victim's code but instead applies precisely-timed pulses to their \textit{own} qubits to intentionally induce errors on a neighboring victim qubit. Such attacks are particularly insidious as they operate below the standard gate-level abstraction and leverage a purely physical phenomenon. There currently lacks a comprehensive framework to rigorously connect these physical interactions to interpretable, logical error models. To date, no study has adequately addressed these crucial questions: \\

\noindent\hypertarget{rq1}{\underline{\textbf{RQ1.}}} Can hardware-level crosstalk be systematically exploited to actively induce targeted logical errors in quantum computations, thereby compromising computational integrity?

\noindent\hypertarget{rq2}{\underline{\textbf{RQ2.}}} What distinct and asymmetric roles do individual adversarial qubits play in structured pulse-level crosstalk attacks?

\noindent\hypertarget{rq3}{\underline{\textbf{RQ3.}}} How robust and stealthy are these pulse-induced crosstalk attacks across various quantum computing protocols and typical hardware parameter variations, such as frequency detuning?\\

\textbf{This paper directly addresses these questions by introducing the first comprehensive, end-to-end pulse-to-circuit characterization framework tailored to active, integrity-focused quantum crosstalk attacks.} Our key contributions include:
\vspace{-0.5em}
\begin{enumerate}
    \item A rigorous, channel-theoretic characterization methodology integrating quantum process tomography (QPT), Kraus decomposition, and a novel isometry-based optimization, effectively translating low-level physical pulse-level interactions into intuitive and interpretable logical error circuits.
    \item Detailed characterization of asymmetric "driver" and "catalyst" qubit roles within structured multi-qubit crosstalk attacks, revealing sophisticated and previously overlooked adversarial dynamics.
    \item Validation demonstrating both the potency and stealth of these attacks across a diverse set of quantum protocols—including state preparation, quantum binary classification tasks, and quantum neural network (QNN) primitives, highlighting their practical implications across quantum computational domains.
    \item Quantitative analysis of the attack’s robustness to realistic hardware imperfections, such as frequency detuning, underscoring the real-world viability and danger posed to current multi-tenant quantum hardware.
\end{enumerate}

\noindent
\textbf{Paper Structure:}
Sec.~\ref{sec:background and threat model} defines our threat model and reviews the essential physical and mathematical formalisms underpinning crosstalk-induced quantum channels. Sec.~\ref{sec:attack framework} outlines the attack framework, including pulse parameters and Hamiltonian modeling methods. Sec.~\ref{sec:crosstalk channel} systematically identifies dominant crosstalk mechanisms, pulse shapes, and demonstrates their protocol-level impact. Sec.~\ref{sec:Attack Impact on Quantum Protocols} introduces our novel channel reconstruction and isometry-based circuit fitting pipeline, clearly linking physical attacks to interpretable logical models. Sec.~\ref{sec:Characterizing the Attack Mechanism} provides comprehensive numerical results that reinforce and extend insights from our logical characterization, including asymmetric roles and attack robustness. Finally, we propose a practical mitigation framework based on observed "attack signatures," discussing the broader implications of our findings for quantum hardware security, and outlining promising directions for future research.

\section{Background and Threat Model}\label{sec:background and threat model}

Quantum computation relies on the precise manipulation of coupled qubit systems, yet this very coupling introduces both essential computational capability and potential vulnerabilities. This section reviews the physical origins of inter-qubit crosstalk in superconducting architectures, sets up the adversarial threat model for multi-tenant systems, and introduces the mathematical formalism of quantum channels, which forms the basis for our later analysis of crosstalk-induced attacks and their characterization.

\subsection{Qubit Coupling and Crosstalk}
In superconducting quantum computers, logical operations are performed by applying precisely-shaped microwave pulses to manipulate the quantum states of individual qubits. While single-qubit gates act on isolated qubits, multi-qubit gates, which are necessary for creating entanglement and executing universal quantum algorithms, depend on engineered interactions between two or more qubits \cite{ketterer2023characterizing}. This inter-qubit coupling is physically realized through elements like capacitive or inductive links, and the resulting interactions are described by specific terms in the system's Hamiltonian, such as longitudinal (ZZ) or transverse (ZX, YX) couplings \cite{zhao2022quantum}. These can be generically written as:

\begin{equation}
    H_{\text{coup}} = J_{ZZ} \, \sigma_z^{(i)} \otimes \sigma_z^{(j)} 
                    + J_{ZX} \, \sigma_z^{(i)} \otimes \sigma_x^{(j)}
                    + J_{YX} \, \sigma_y^{(i)} \otimes \sigma_x^{(j)}
    \label{eq:generic_crosstalk}
\end{equation}
where $J_{ZZ}$, $J_{ZX}$, and $J_{YX}$ are coupling strengths, and $\sigma_\alpha^{(i)}$ denotes the Pauli operator acting on qubit $i$.

The engineered couplings, while essential for multi-qubit operations, also create pathways for adversarial exploitation, particularly in multi-tenant environments where qubits may be assigned to different users. This dual nature of coupling, thus act as both: a computational necessity and a security liability \cite{ketterer2023characterizing}. 

Crosstalk emerges as an unavoidable consequence of designing interconnected qubit systems, representing any unintended or parasitic interaction between qubits or their control lines. From a physics standpoint, crosstalk manifests as unwanted, always-on Hamiltonian coupling terms that are not part of the intended gate operation. The most prevalent of these is static ZZ interaction, which can lead to correlated, coherent errors across processors \cite{mundada2019suppression}. These parasitic effects degrade computational fidelity and are a significant obstacle to scaling quantum hardware, prompting the development of mitigation techniques like dynamical decoupling \cite{tripathi2022suppression}. While often treated as a source of computational noise to be mitigated, from a security perspective, these parasitic couplings constitute a physical-layer side channel. The same Hamiltonian terms that cause noise can be deliberately modulated and exploited \cite{zhao2022quantum}. As described in the threat model of this paper, an adversary with control over their own qubits can inject specifically engineered pulses to leverage these crosstalk pathways. This allows them to induce targeted logical errors on a physically adjacent victim qubit without requiring any direct access, transforming crosstalk from a simple fidelity issue into a potent attack vector.

\subsection{Threat Model}
\begin{figure}[!htb]
    \centering
    \includegraphics[width=0.8\columnwidth]{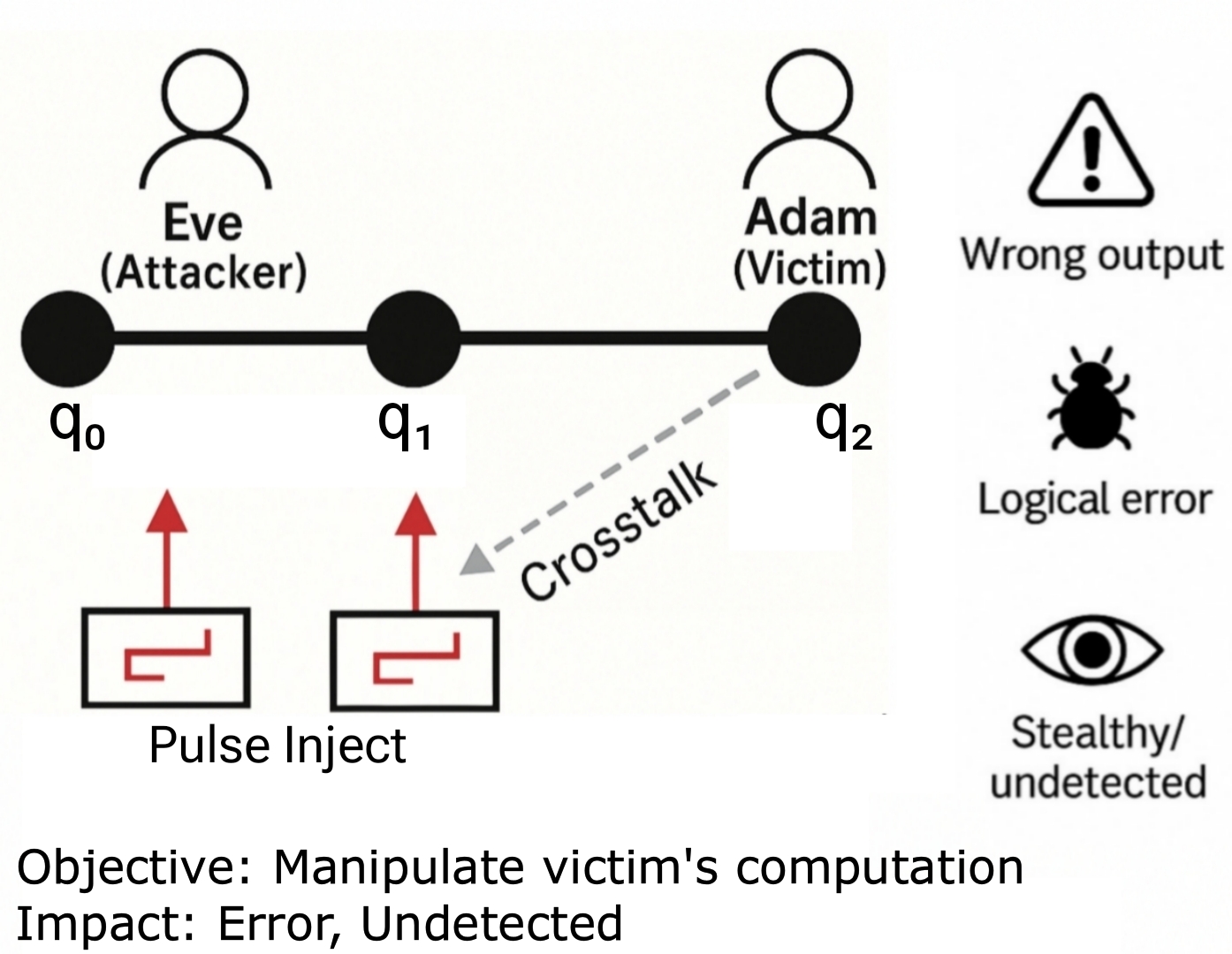}
    \caption{Conceptual illustration of a crosstalk attack. An attacker (Eve) injects malicious pulses targeting qubits $q_0$ and $q_1$. This generates crosstalk noise that affects a physically adjacent qubit, $q_2$, used by the victim (Adam), leading to stealthy and undetected logical errors.}
    \label{fig:crosstalk_attack}
\end{figure}

We define a threat model in a multi-tenant environment, conceptually illustrated in Figure~\ref{fig:crosstalk_attack}, involving an attacker (Eve) and a victim (Adam) sharing a quantum processor with a linear qubit topology: $q_0-q_1-q_2$.

Eve has full pulse-level control over her qubits ($q_0, q_1$), which are physically adjacent to Adam's qubit ($q_2$). \\

\noindent \textbf{\underline {Attacker Objective and Impact}}
Eve's primary objective is to exploit native hardware crosstalk (e.g., ZX or YX coupling) to induce unintended quantum operations on Adam's qubit, $q_2$. By injecting tailored pulses into her own qubits, she aims to corrupt the outcome of the victim’s computation without any direct access to, or knowledge of, their circuit or data.

The intended impact of this attack is threefold:
\begin{itemize}
    \item \textbf{Computational Degradation:} The attack degrades the victim's computation, causing wrong outputs or reduced fidelity.
    \item \textbf{Silent Errors:} It can introduce subtle logical errors that go undetected by standard verification routines.
    \item \textbf{Stealth:} The effects are designed to mimic device calibration drift or random noise, making the attack difficult to attribute and distinguish from normal hardware imperfections.
\end{itemize}

\vspace{1em}
\noindent \textbf{\underline{Capabilities and Limitations}}
The model assumes the following capabilities:
\begin{itemize}
    \item \textbf{Eve (Attacker):} Possesses complete, low-level software or physical access to her assigned qubits ($q_0, q_1$), allowing for precise manipulation of microwave pulse parameters (shape, amplitude, frequency). However, she lacks direct access to $q_2$ and has no knowledge of the specific algorithm Adam is running.
    \item \textbf{Adam (Victim):} Operates $q_2$ using standard gate-based commands and is entirely unaware of Eve's malicious activity, assuming normal hardware functionality.
\end{itemize}

\subsection{Quantum Channels and Isometry Freedom}
\label{sec:channels_isometry}

Any closed quantum system evolves unitarily: $\rho \mapsto U \rho U^\dagger$. In practical quantum hardware, however, we are often concerned with the effective evolution of a subsystem—such as a victim’s qubit, when it is coupled to auxiliary qubits (e.g., those controlled by an adversary). The dynamics of the subsystem are described by a quantum channel, i.e., a completely positive, trace-preserving (CPTP) map, obtained by tracing out the auxiliary degrees of freedom:
\begin{equation}
    \mathcal{E}(\rho) = \mathrm{Tr}_{\text{aux}} \left[ U\, (\rho \otimes \tau_{\text{aux}})\, U^\dagger \right]
    \label{Eq:Kraus}
\end{equation}
where $\tau_{\text{aux}}$ is the initial state of the auxiliary (potentially adversarial) qubits.

Even if victim's state start as pure, i.e., $\rho = \ket{\psi}\bra{\psi}$, due to the CPTP map, the resulting state $\rho'=\mathcal{E}(\rho)$ may convert into a mixed state (i.e., $Tr[\rho'^2]<1$). 

Now any quantum channel defined by Eq.~\ref{Eq:Kraus} admits a Kraus decomposition \cite{nielsen2010quantum}:
\begin{equation}
    \mathcal{E}(\rho) = \sum_{j=1}^{r} K_j \rho K_j^\dagger, \qquad \sum_{j=1}^{r} K_j^\dagger K_j = I
    \label{Eq:KrausDecomp}
\end{equation}
where $r$ (the number of Kraus operators) depends on the size of the traced-out auxiliary system. Importantly, this decomposition is not unique: any two sets of Kraus operators $\{K_j\}$ and $\{K'_j\}$ that are related by a unitary (isometry) transformation of size $r$ describe the same quantum channel:
\begin{equation}
    K'_j = \sum_{i=1}^{r} U_{ji} K_i
    \label{Eq:Isometry}
\end{equation}
where $U^\dagger U = I$. This isometry gauge freedom is central to both quantum process tomography and our model fitting procedure, as it must be accounted for when comparing theoretical and experimental channels.

To circumvent this gauge ambiguity, one may instead use the Choi matrix $C_{\mathcal{E}}$, a unique and basis-independent representation \cite{wood2011tensor} of the channel, defined as:
\begin{equation}
    C_{\mathcal{E}} = \sum_{m,n} \ket{m}\bra{n} \otimes \mathcal{E}(\ket{m}\bra{n})
    \label{Eq:Choi}
\end{equation}
The Choi matrix uniquely characterizes the action of the quantum channel and provides a systematic route to reconstructing all possible Kraus representations via its eigen-decomposition. In this work, the Choi matrix formalism allows us to compare experiment and model directly and unambiguously, as different Kraus sets with the same Choi matrix represent the same physical channel.

This mathematical machinery underpins our approach to process tomography, isometry fitting, and quantitative attack analysis throughout the remainder of the paper.

\section{Attack Framework}\label{sec:attack framework}
This section details framework used to model and execute the pulse-level crosstalk attack. We first define two primary strategic approaches an adversary can employ based on timing of the pulse injection relative to the victim's operations. We then present the physical model of the attack, including the specific pulse parameters under the adversary's control and the time-dependent Hamiltonian that governs the system's evolution in the rotating frame.

\subsection{Attack Strategies and Timing}
The efficacy of the crosstalk attack is highly dependent on the timing of the malicious pulse injection relative to the victim's operations. Based on this, the adversary, Eve, can employ two distinct strategies:

\begin{itemize}
    \item \textbf{Attacker-First:} Eve preemptively applies malicious pulses to $q_0$ and $q_1$ \textit{before} the victim, Adam, initializes or operates on $q_2$. This approach aims to corrupt the effective initial state of the victim's qubit, thereby influencing all subsequent operations. As our results will show, this is the most damaging scenario.
    \item \textbf{Victim-First:} Eve injects her pulses \textit{after} Adam has already prepared or operated on the state of $q_2$. This strategy aims to disrupt the prepared state through crosstalk effects. It is generally less impactful than the attacker-first approach but can be stealthier.
\end{itemize}

\subsection{Physical Attack Model}
Eve’s attack relies on carefully designed microwave pulses applied to $q_0$ and $q_1$ to induce crosstalk impacting $q_2$. In \textbf{rotating frame}, the key pulse parameters include:

\begin{itemize}

\item \textit{Pulse Amplitude ($A$):} The strength of the applied microwave pulse.

\item \textit{Frequency Detuning ($\delta$):} The offset between the pulse frequency and the qubit's resonant frequency.

\item \textit{Chirp Rate ($c$):} The rate of frequency change over time in chirped pulses.

\item \textit{DRAG Parameter ($\alpha$):} A correction factor in DRAG pulses to mitigate higher-order effects.

\item \textit{Pulse Width ($\sigma$):} The duration parameter for Gaussian and DRAG pulses.

\item \textit{Pulse Shape:} The temporal profile of the pulse, with options detailed in Table~\ref{tab:pulse_shapes}.

\end{itemize}

The mathematical forms of the pulse shapes analyzed in this work are detailed in Table~\ref{tab:pulse_shapes}.

\begin{table}[!htb]
\caption{Mathematical definitions of pulse shapes employed in simulations.}
\label{tab:pulse_shapes}
\setlength\tabcolsep{3pt}
\scriptsize
\resizebox{0.45\textwidth}{!}{
\setlength\extrarowheight{0pt}
\begin{tabular}{ll}
\toprule
Shape & Function $f(t)$ \\
\midrule
Cosine & $A \cos(\delta t)$ \\
Gaussian & $A \exp\left(-\frac{(t-0.5)^2}{2\sigma^2}\right)$ \\
Square & $A$ if $0.3 \le t \le 0.7$, else $0$ \\
Chirp & $A \cos\left[(\delta + c \cdot t)t\right]$ \\
DRAG & $A\left[ \exp\left(-\frac{(t-0.5)^2}{2\sigma^2}\right) - \alpha\frac{t-0.5}{\sigma^2}\exp\left(-\frac{(t-0.5)^2}{2\sigma^2}\right)\right]$ \\
\bottomrule
\end{tabular}}
\end{table}

\subsection{System Dynamics and Simulation}

To accurately simulate the physical interaction of the crosstalk attack, we model the system's dynamics using a time-dependent Hamiltonian in the \emph{rotating frame}. This is a standard and computationally efficient technique in simulating driven quantum systems, as it removes the fast-oscillating terms associated with the qubits' natural frequencies. This allows the simulation to focus on the slower dynamics induced by the control pulses and coupling terms, which are the basis of the logical operations and the attack itself \cite{wei2022hamiltonian, balewski2025first}.
The total Hamiltonian is the sum of a static component representing the physical crosstalk and a time-dependent component representing the adversary's drive pulses:
\begin{equation}
    H(t) = H_{\text{coupling}} + H_{\text{drive}}(t).
\end{equation}

\noindent\underline{\textbf{Crosstalk Hamiltonian}}
The static, time-independent term, $H_{\text{coupling}}$, models the persistent, always-on parasitic interaction between adjacent qubits. This unwanted coupling is an inherent property of the physical hardware architecture. We model it as:
\begin{equation}
    H_{\text{coupling}} = J_{01} \left(\sigma^{(0)} \otimes \sigma^{(1)} \otimes I\right) + J_{12}\left(I \otimes \sigma^{(1)} \otimes \sigma^{(2)}\right)
\end{equation}
where $J_{01}$ and $J_{12}$ are the scalar coupling strengths, and the specific Pauli operators ($\sigma^{(i)}$) for the interaction (e.g., ZX or YX coupling) are varied in our analysis.

\noindent\underline{\textbf{Drive Hamiltonian}}
The time-dependent term, $H_{\text{drive}}(t)$, represents the external microwave pulses that Eve actively controls and injects into her qubits. This drive is modeled as:
\begin{equation}
    H_{\text{drive}}(t) = A_0 f_{0}(t)\sigma_{x}^{(0)} \otimes I \otimes I + A_1 f_{1}(t) I \otimes \sigma_{x}^{(1)} \otimes I
\end{equation}
where $A_0$ and $A_1$ are the respective pulse amplitudes, and $f_0(t)$ and $f_1(t)$ are the normalized, time-dependent pulse shapes as defined in Table~\ref{tab:pulse_shapes}.

\noindent\underline{\textbf{Numerical Simulation}}
The time evolution of the three-qubit quantum state, $\ket{\psi(t)}$, under adversarial pulse-level attacks is governed by the time-dependent Schrödinger equation:
\begin{equation}
    i\hbar \frac{\partial}{\partial t}\ket{\psi(t)} = H(t)\ket{\psi(t)}
\end{equation}
where $H(t)$ is the total Hamiltonian incorporating both static coupling and time-dependent drive terms.
Given an initial state $\ket{\psi(0)}$, we numerically integrate this equation to obtain the final state $\ket{\psi(1)}$ after application of the adversarial pulses.

The simulation is implemented using the QuTiP library’s `mesolve` master equation solver \cite{li2022pulse}. To model a closed quantum system (no decoherence), the list of collapse operators is left empty, ensuring purely unitary evolution. The time evolution is performed over a normalized interval $t \in [0, 1]$, discretized into 50 uniform time steps. 
We then explicitly construct the final density matrix $\rho_{\text{final}}$ of the entire three-qubit register. We then obtain the reduced density matrix for the victim’s qubit ($q_2$) by tracing out the adversarial qubits:
$\rho_{q_2} = \mathrm{Tr}_{q_0, q_1}\left[\rho_{\text{final}}\right]
$

All simulation parameters, including pulse shapes, coupling types, amplitudes, and detunings, are systematically varied as described in previous subsection to identify high-influence crosstalk channels and quantify their impact on victim’s logical state.

\section{Identifying Dominant Crosstalk Channels}\label{sec:crosstalk channel}

In order to effectively design and mitigate crosstalk-based attacks, it is critical to first understand which coupling types and pulse shapes are most capable of inducing significant logical errors on a victim qubit. Motivated by this necessity, we systematically characterized the influence of various crosstalk channels by simulating the system response under different configurations. To empirically identify the dominant pathways for logical state corruption, we fixed a reference cosine pulse on the "driver" qubit ($q_1$) and systematically varied the drive on the "catalyst" qubit ($q_0$). The induced effect on the victim qubit ($q_2$) was then measured without any direct control or prior knowledge of the victim's operations, quantifying the indirect coupling mediated through the interaction Hamiltonian.

The key metric is the $L_2$ norm between the outcome distributions with and without attack:
\begin{equation}
\|\Delta p\|_2 = \left[ \sum_i \left( p_i^\mathrm{(attack)} - p_i^\mathrm{(ideal)} \right)^2 \right]^{1/2}
\end{equation}
where $p_i$ are measurement probabilities for $q_2$.

\begin{table}[!htb]
\caption{Dominant crosstalk coupling types and their quantified influence ($\|\Delta p\|_2$) for key pulse shapes, with reference configuration: $J_{01}=J_{12}$, $q_1$ cosine drive.}
\label{tab:coupling_scan}
\begin{tabular}{lcc}
\toprule
Coupling Type ($\sigma_a \otimes \sigma_b$) & Pulse Shape & Influence Norm ($\|\Delta p\|_2$) \\
\midrule
Y $\otimes$ X & Chirp   & 0.0108 \\
Y $\otimes$ X & Cosine  & 0.0108 \\
Z $\otimes$ X & Chirp   & 0.0073 \\
Z $\otimes$ X & Cosine  & 0.0073 \\
Y $\otimes$ X & Square  & 0.0037 \\
Y $\otimes$ X & DRAG    & 0.0027 \\
Y $\otimes$ X & Gaussian& 0.0016 \\
Z $\otimes$ X & DRAG    & 0.0014 \\
Z $\otimes$ X & Square  & 0.0012 \\
Z $\otimes$ X & Gaussian& 0.0004 \\
\bottomrule
\end{tabular}
\end{table}

The coupling scan summarized in Table~\ref{tab:coupling_scan} demonstrates that the most significant logical errors are generated by off-diagonal crosstalk terms, specifically $Y\otimes X$ and $Z\otimes X$, in combination with chirp or cosine pulse shapes. Other configurations have an influence norm below $10^{-4}$, meaning $q_0$ in those cases can barely affect $q_2$. These findings justify our focus on these high-impact channels for all subsequent security experiments.

While the direct neighbor $q_1$ will always have a dominant crosstalk effect on $q_2$, the characterization of the mediated attack from $q_0$ reveals a more profound security vulnerability. Two users' qubits separated by an unused qubit can also face security threats. We demonstrate that even when an attacker (Eve, on $q_0$) and a victim (Adam, on $q_2$) are not direct neighbors, Eve can still leverage the intermediate qubit ($q_1$) as a conduit to execute a targeted, stealthy integrity attack. This "attack-through-a-neighbor" vector underscores a far more complex and dangerous threat landscape than previously considered, with significant implications for secure resource allocation and scheduling on future quantum hardware.

\section{Attack Impact on Quantum Protocols}\label{sec:Attack Impact on Quantum Protocols}

In this section, we address \hyperlink{rq1}{\textbf{RQ1}} by explicitly demonstrating how hardware-level crosstalk can systematically induce targeted logical errors that compromise computational integrity, by simulating its effect on several single-qubit quantum protocols (Sections~\ref{sec:coin_flip}-\ref{sec:sqqnn}). We also partially address \hyperlink{rq3}{\textbf{RQ3}} by highlighting protocol-dependent vulnerability and stealth characteristics of these attacks (Section~\ref{sec:security_implications}).

\begin{figure}[!htb]
\centering
\scalebox{1.0}{
\Qcircuit @C=1.0em @R=0.5em @!R {
\nghost{q : } & \lstick{q : } & \control\qw & \gate{\mathrm{V}(\lambda)} & \control\qw & \meter & \qw & \qw \\
\nghost{} & & \raisebox{0.3em}{a} & & \raisebox{0.3em}{b} & & & \\
\nghost{c : } & \lstick{c : } & \cw & \cw & \cw & \dstick{_{0}} \cw \ar @{<=} [-2,0] & \cw & \cw \\
\nghost{} & & & & & & & \\
}
}

\vspace{-2em}
 \caption{General experimental circuit to validate adversarial impact. The attack can be injected at point 'a' (attacker-first) or 'b' (victim-first) relative to the victim's operation $V(\lambda)$.}

 \label{fig:exp_setup}
\end{figure}
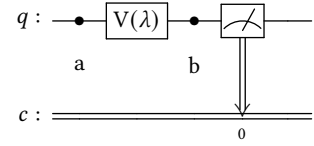

The general experimental setup is shown in Figure~\ref{fig:exp_setup}, where a crosstalk attack, implemented with a cosine pulse, is injected either before the victim's core operation $V(\lambda)$ (the "attacker-first" strategy at point 'a') or after it (the "victim-first" strategy at point 'b'). This allows us to assess how the attack's timing relative to the victim's computation influences its effectiveness.

\subsection{Biased Quantum Coin Flip}\label{sec:coin_flip}
We first test the attack's ability to corrupt a simple state preparation task. In this protocol, the victim's qubit is prepared in the state $\ket{\psi(\lambda)} = \cos\lambda\ket{0} + \sin\lambda\ket{1}$.

To prepare the victim circuit as shown in Figure \ref{fig:exp_setup}, we put:
\begin{equation}
V(\lambda)= e^{-i\lambda Y} = \left[ \begin{smallmatrix}
    cos(\lambda) & -sin(\lambda) \\
    sin(\lambda) & cos(\lambda)
\end{smallmatrix}  \right]
\end{equation}
where $Y$ is the Pauli-Y matrix. The ideal probability of measuring $\ket{1}$ is $P(\ket{1})=\sin^2\lambda$. The attack's impact is demonstrated under two distinct parameter configurations. 

\begin{figure}[!htb]
  \centering
  \includegraphics[width=0.49\textwidth]{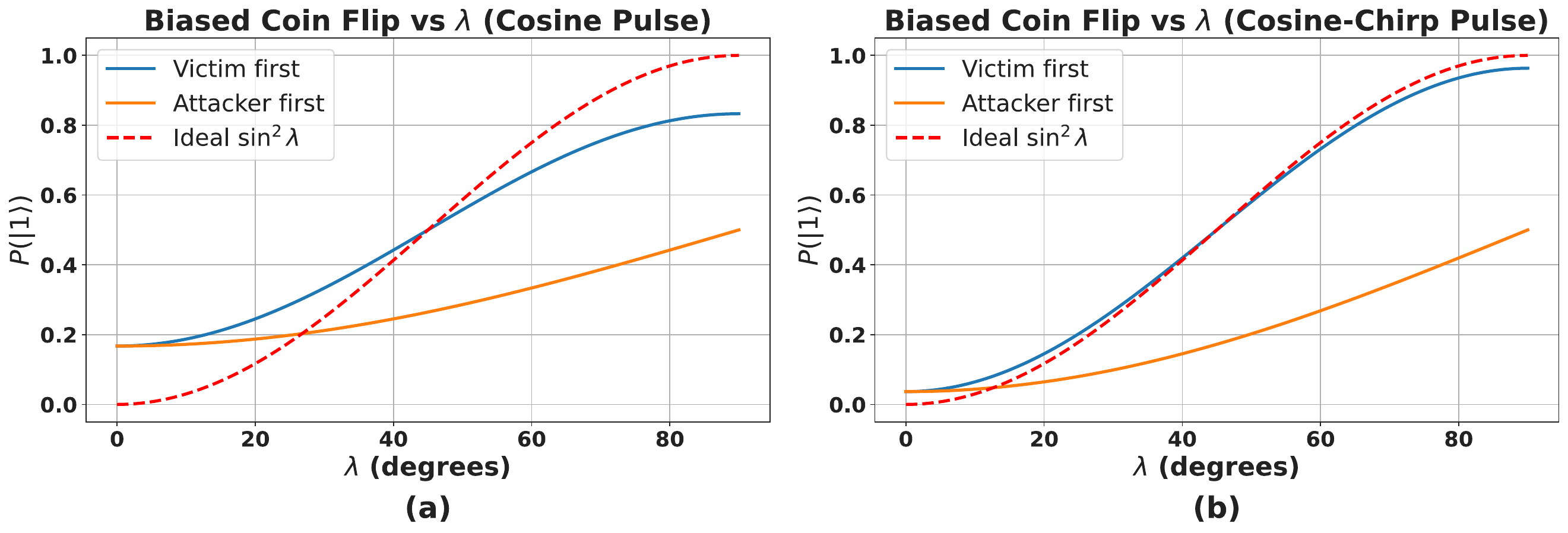}
  \caption{
    Impact of pulse-level crosstalk attacks on the Biased Quantum Coin Flip protocol. 
    \textbf{(a)} Standard configuration: Cosine–Cosine pulses with moderate amplitudes ($A_0 = A_1 = 0.5$). 
    \textbf{(b)} Aggressive configuration: Cosine–Chirp pulse combination with maximal amplitudes ($A_0 = A_1 = 1.0$).
    The attack alters the measurement statistics as a function of the input rotation angle $\lambda$, highlighting the increased bias introduced by more aggressive pulse settings.
  }
  \label{fig:coinflip_default}
\end{figure}

First, we simulate a default attack scenario using cosine pulses on both adversarial qubits with moderate amplitudes ($A_0=A_1=0.5$) and a coupling strength of $J_{01}=J_{12}=0.5$. As shown in Figure~\ref{fig:coinflip_default} (a), this configuration successfully biases the outcome. The attacker-first strategy causes a significant and predictable deviation from the ideal curve, while the victim-first strategy has a more subtle effect, becoming nearly invisible at certain angles, which highlights its potential for stealth \cite{shubha2025qsec}.

Next, to demonstrate a maximal-strength attack, we increase the amplitudes to their maximum ($A_0=A_1=1.0$) and utilize a more aggressive Cosine-Chirp pulse shape combination. The results in Figure~\ref{fig:coinflip_default} (b) show a much more pronounced deviation, especially for the attacker-first strategy. Together, these experiments demonstrate that an adversary can tune the attack's intensity and that even non-identical pulse shapes can be combined to effectively compromise the integrity of quantum state preparation.

\subsection{XOR Quantum Classifier}\label{sec:xor}
\begin{figure}[!htb]
\centering
\scalebox{1.0}{
\Qcircuit @C=1.0em @R=0.2em @!R {
\lstick{q : } & \gate{\ket{0}} & \gate{\mathrm{H}} & \gate{\mathrm{R_Z}((2x_1 - 1)\frac{\pi}{2})} & \gate{\mathrm{R_X}((2x_2 - 1)\frac{\pi}{2})} & \meter & \qw \\
}
}
\caption{Quantum circuit for the XOR classification task. Classical inputs $x_1, x_2$ are encoded into rotation angles.}
\label{fig:xor_circuit}
\vspace{-1em}
\end{figure}
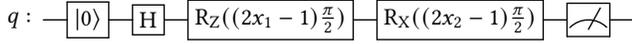

Next, we analyze the attack's impact on a simple but non-trivial logical circuit: a single-qubit XOR classifier \cite{grossu2021single}. This circuit, shown in Figure~\ref{fig:xor_circuit}, encodes two classical bits, $x_1, x_2 \in \{0, 1\}$, into the rotation angles of sequential gates applied to a qubit initialized in the $\ket{0}$ state.
The output of this sequence, $\ket{\psi_{\text{out}}} = U_{\text{XOR}}(x_1, x_2)\ket{0}$, results in a final state that corresponds to the logical XOR of the inputs, $\ket{x_1 \oplus x_2}$, up to a global phase. The final measurement in the computational basis therefore deterministically classifies the XOR result.

However, when there is an attack, the map becomes CPTP, and the resultant state is a mixed state $\rho_{x_1,x_2}$.
Now, to quantify the attack's impact, we measure the worse case error over all four possible input pairs, which we defined as:

\begin{equation}
\Delta_{\text{max}} = \left| 1 - \min_{(x_1,x_2) \in \{0,1\}^2} \text{max} \left[P_{x_1,x_2}(\ket{0}), P_{x_1,x_2}(\ket{1})\right ] \right|
\end{equation}

Here, $P_{x_1,x_2}(\ket{k})= \bra{k}\rho_{x_1, x_2}\ket{k}$. \\

\begin{figure}[h!]
  \centering
  \includegraphics[width=0.95\linewidth]{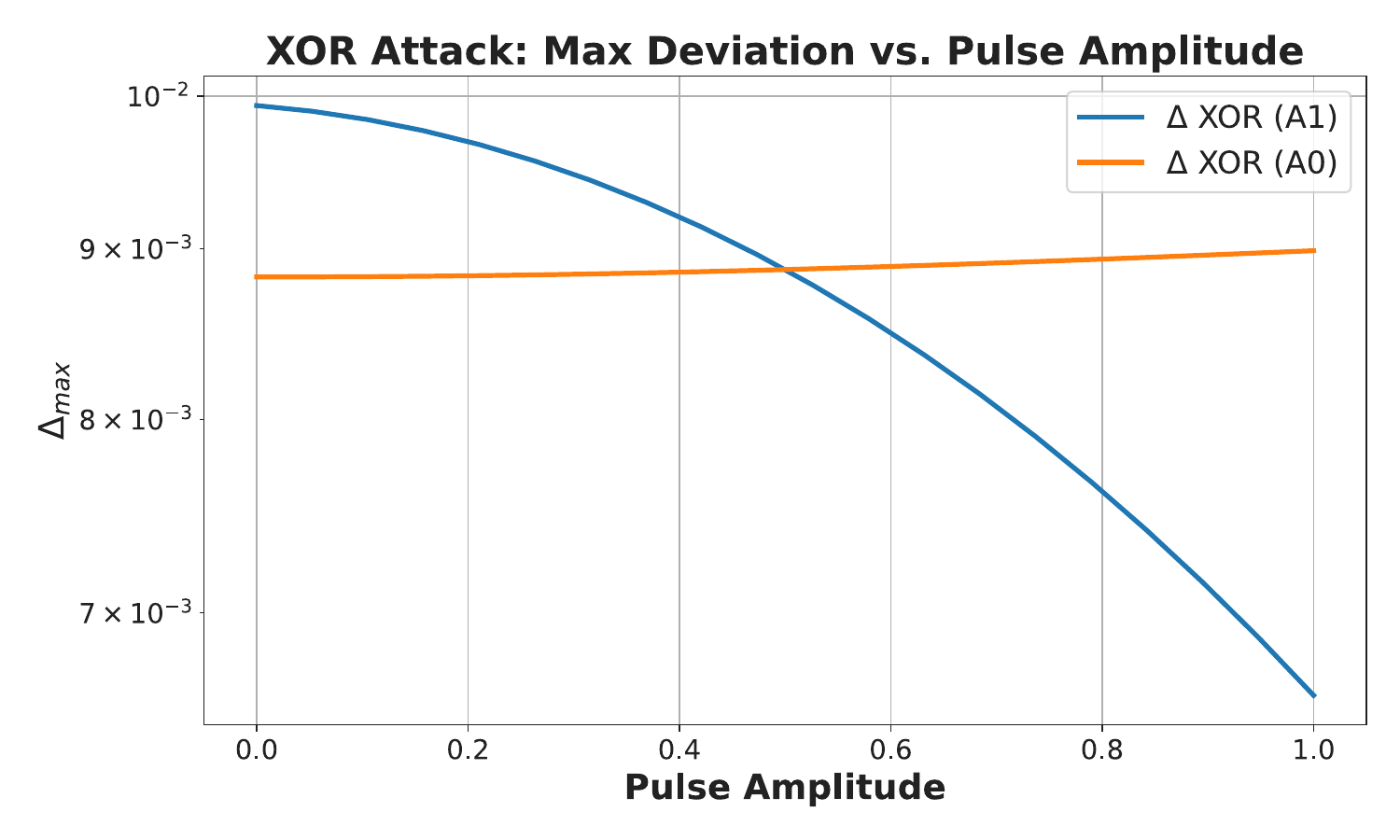}
  \vspace{-0.5em}
  \caption{Maximum deviation in output statistics for the XOR quantum classifier as a function of pulse amplitude. The classifier's robustness is evident, with only minor deviations even at high attack strength.}
  \label{fig:xor_results}
\end{figure}

Remarkably, we found this protocol to be highly robust against our crosstalk attack for both attack scenarios. As shown in Figure~\ref{fig:xor_results}, the deviation in output probabilities is minimal (on the order of $10^{-2}$) even under maximal attack amplitudes \cite{shubha2025qsec}. The reason for this resilience lies in circuit's design. The small, coherent rotation induced by crosstalk attack is insufficient to push final state across a decision boundary, leaving logical outcome unchanged. The near-abelian structure of this circuit with respect to induced error channel makes it resilient, a subject that warrants further study.

An interesting secondary observation from Figure~\ref{fig:xor_results} is the subtle, counter-intuitive effect of the two adversarial qubits. While the overall error is small, increasing the amplitude of the "driver" qubit ($A_1$) actually helps to slightly decrease the error, while increasing the "catalyst" qubit's amplitude ($A_0$) slightly increases it. This suggests that even when an attack is largely ineffective, the driver and catalyst roles are still present, creating a complex interplay. An advanced adversary could potentially leverage this effect to fine-tune their attack strategy, balancing different pulse amplitudes to remain more stealthy while still inducing a small, targeted bias.

\vspace{-0.5em}

\subsection{Single-Qubit Quantum Neural Network (SQQNN)}\label{sec:sqqnn}
In contrast to the XOR classifier, our results show that a Single-Qubit Quantum Neural Network \cite{souza2024regression} is acutely vulnerable to the crosstalk attack. We trained an SQQNN to perform binary classification on the Iris dataset, where the model learns optimal rotation angles to define a sensitive decision boundary.

The SQQNN model begins with classical pre-processing. Given input data $\mathbf{x}_i \in \mathbb{R}^p$ and target labels $y_i \in \{-1, 1\}$, we first construct polynomial features to create a richer representation:
\begin{equation}\small
    \mathbf{x}_{\text{poly},i} = [1, x_{i1}, x_{i2}, \ldots, x_{ip}, x_{i1}^2, x_{i2}^2, \ldots, x_{ip}^2, \ldots, x_{i1}^K, x_{i2}^K, \ldots, x_{ip}^K]
\end{equation}

where $K$ is the polynomial degree. To prepare the target labels for the model and avoid singularities in the subsequent $arctanh$ function, we apply a transformation with a small regularization parameter $\epsilon = 10^{-16}$:

\begin{equation}
    y'_i = \begin{cases}
1 - \epsilon & \text{if } y_i = 1 \\
-1 + \epsilon & \text{if } y_i = -1
\end{cases}
\end{equation}

The fully transformed target vector for training is then $\mathbf{Y} = [\text{arctanh}(y'_1), \ldots, \text{arctanh}(y'_n)]^T$. 

The model's coefficients, $\mathbf{S} = [c_0, c_{11}, \ldots, c_{Kp}]^T$, are determined by solving a linear least squares problem:
\begin{equation}
    \mathbf{S} = (\mathbf{X}^T \mathbf{X})^{-1} \mathbf{X}^T \mathbf{Y}
\end{equation}

where $\mathbf{X}$ is the design matrix composed of the polynomial features.

\vspace{\baselineskip}
\noindent\textbf{Quantum Circuit Implementation:}
The trained classical coefficients are then used to parameterize a quantum circuit. For each test sample $\mathbf{x}_{\text{poly},i}$, we compute a value $z_i$:

\begin{equation}
    z_i = \mathbf{S}^T \mathbf{x}_{\text{poly},i} = c_0 + \sum_{k=1}^{K} \sum_{j=1}^{p} c_{kj} x_{ij}^k
\end{equation}

This value is mapped to a rotation angle $\beta_i = \arccos(\tanh(z_i))$. Adam's circuit then applies this rotation to an initial state $|0\rangle$ using a Y-rotation gate, $R_y(\beta) = e^{-i\beta\sigma_y/2}$:

\begin{equation}
   | \psi_{\text{final}}\rangle = R_y(\beta_i) |0\rangle
\end{equation}

The final prediction is obtained by measuring the expectation value of the Pauli-Z operator, and taking its sign:
\begin{equation}
    \langle\sigma_z\rangle = \langle\psi_{\text{final}}|\sigma_z|\psi_{\text{final}}\rangle, \quad \hat{y}_i = \text{sign}(\langle\sigma_z\rangle)
\end{equation}

\vspace{\baselineskip}
\noindent\textbf{Performance Evaluation and Vulnerability Analysis:}
To evaluate the model's performance under attack, we use the accuracy metric, which is the fraction of correctly classified test samples:
\begin{equation}
\text{Accuracy} = \frac{1}{n_{\text{test}}} \sum_{i=1}^{n_{\text{test}}} \mathbb{I}[\hat{y}_i = y_i]    
\end{equation}

\vspace{-6pt}
\begin{figure}[!htb]
  \centering
  \includegraphics[width=0.9\linewidth]{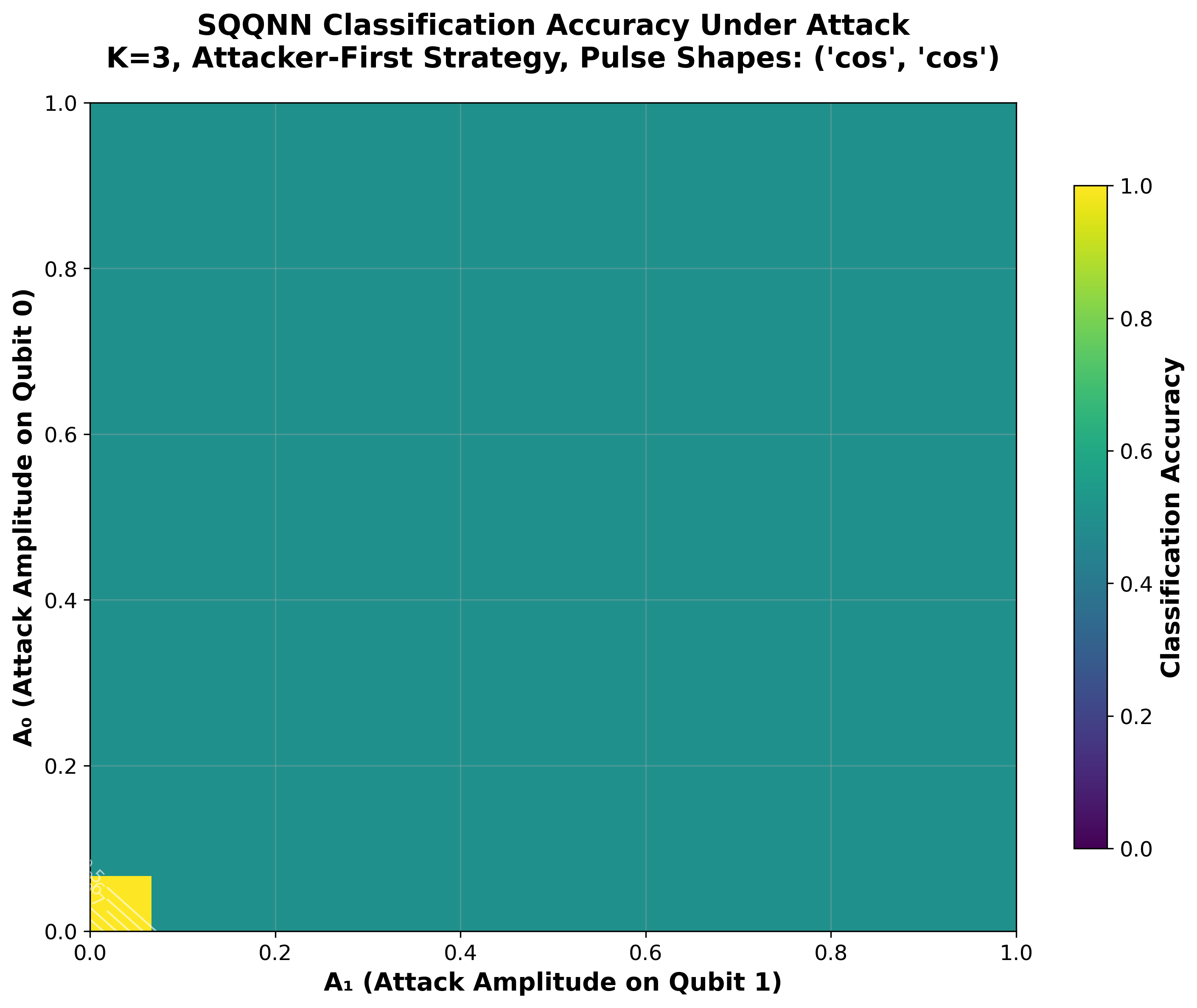}
  \vspace{-6pt}
  \caption{Classification accuracy heatmap of a Single-Qubit QNN under attack. The accuracy drops to 0.5 for any possible attack, confirming the vulnerability of QML models to this attack vector.}
  \label{fig:sqqnn}
\end{figure}

The effectiveness of an attack is measured by the drop in this accuracy: $\Delta_{\text{accuracy}} = \text{Accuracy}_{\text{no attack}} - \text{Accuracy}_{\text{attack}}$ \\
Our simulations reveal that the SQQNN is acutely vulnerable to the crosstalk attack, but this vulnerability is highly dependent on both the attack strategy and the nature of the model's training. While the victim-first attack causes only a minor negligible drop in performance, the \textbf{attacker-first strategy is catastrophic} as shown in Figure~\ref{fig:sqqnn}. As the attack amplitude increases, the classification accuracy in Attacker-First strategy plummets from nearly 1.0 (perfect) to 0.5 (equivalent to random guessing), rendering the trained model completely useless.

This extreme vulnerability stems directly from the model's single-step, analytical training process. Unlike iterative machine learning models that might gradually adapt to noise, this SQQNN determines its weights, $\mathbf{S}$, in a single calculation by solving a linear least squares problem. This makes it exceptionally brittle. The attacker-first strategy exploits this perfectly: by applying a small, coherent rotation, it corrupts the initial state \textit{before} the crucial feature-to-angle mapping occurs. This effectively poisons the entire dataset from the model's perspective, causing the one-shot weight calculation to produce a completely incorrect function. In contrast, the victim-first attack occurs \textit{after} the core classification rotation has already been applied based on the correct state, resulting in only a minor final perturbation that is insufficient to cause a misclassification. This highlights a \textbf{critical security concern for non-iterative QML models}, where subtle physical-layer errors can completely and irrecoverably invalidate the entire trained network.

\vspace{-15pt}
\subsection{Implications for Hardware Security}\label{sec:security_implications}

Our protocol-level demonstrations of crosstalk attacks highlight several important implications for the security of multi-tenant quantum hardware.

\textbf{1. Protocol-dependent vulnerability: }
A central insight from our experiments is that vulnerability to pulse-level crosstalk attacks is not uniform, but depends strongly on the protocol. For example, protocols that rely on precise initial state preparation, such as quantum coin flip and SQQNN , are significantly more vulnerable when the attacker intervenes during the state initialization (the ``attacker first'' scenario). In these cases, a well-timed, physically plausible pulse from an adversarial tenant can deterministically bias the victim’s computation, as demonstrated in Figures~\ref{fig:coinflip_default} and~\ref{fig:sqqnn}. In contrast, in protocols like the XOR classifier, the logical structure (especially the near-commutation between the attack and the protocol’s native gates) can make both attack scenarios nearly equivalent, substantially mitigating the impact of crosstalk on the measured outputs.

\textbf{2. Stealthiness and attack adaptability: }
Our results also reveal the stealthy nature of these attacks. By carefully choosing pulse parameters and attack timing, an adversary can ensure that the induced errors are either hidden within the protocol's natural noise floor or statistically indistinguishable from ideal outputs in certain parameter regimes. For example, in the unbiased quantum coin flip protocol, an attacker using the victim-first strategy can only bias the outcome for non-maximally biased coins, while remaining undetectable for perfectly fair coins ($\lambda = 45^\circ$). Similarly, in the robust XOR protocol, the impact of both pre- and post-attack strategies is minimal, with errors at or below the device's intrinsic noise level. This stealth property allows attackers to evade detection and selectively target only the most susceptible computations.

\textbf{3. Non-local attack pathways :}
A notable implication is the demonstration of non-local, indirect attack pathways. Even when the attacker (Eve, controlling $q_0$) is not a direct neighbor of the victim (Adam, operating on $q_2$), the adversary can leverage an intermediate qubit ($q_1$) as a conduit. By applying suitably shaped and timed pulses to $q_0$ and $q_1$, the attacker can induce targeted, stealthy integrity violations on $q_2$. This ``attack-through-a-neighbor'' pathway shows that crosstalk vulnerabilities can extend beyond direct couplings, and that the physical topology of the device is a critical security consideration.
These findings are especially relevant for quantum cloud services, where multiple users may be allocated physically adjacent qubits on shared hardware. The ability of an adversarial user to induce protocol-dependent, stealthy errors, potentially even across non-directly coupled qubits, raises new challenges for ensuring isolation and trust. Effective countermeasures will require awareness of protocol structure, physical qubit mapping, and possible indirect crosstalk pathways.

\section{Characterizing the Attack Mechanism}\label{sec:Characterizing the Attack Mechanism}

In order to bridge the gap between the physical, pulse-level attack and an interpretable, logical-level error model, we developed a two-stage characterization framework. 

First, we use Quantum Process Tomography (QPT) to create a complete, model-independent description of the error channel. Second, we introduce a novel isometry-based optimization procedure to approximate this abstract channel with an intuitive, yet powerful, parameterized quantum circuit.

\subsection{Quantum Process Tomography (QPT)}
Quantum Process Tomography enables the complete experimental characterization of an unknown quantum process, $\mathcal{E}$, without any assumptions about its underlying physical nature. This is especially crucial for hardware security, as adversarial pulse-level attacks may induce complex error channels that cannot be captured by simple gate fidelity measurements or error rates alone. The goal is to reconstruct the process matrix, $\chi$, which fully defines the quantum map \cite{giarmatzi2023multi} acting on an arbitrary input density matrix $\rho_{in}$:
\begin{equation}
\rho_{out} = \mathcal{E}(\rho_{in}) = \sum_{m,n=0}^{d^2-1} \chi_{mn} E_m \rho_{in} E_n^\dagger
\end{equation}
where $d=2$ for a single qubit, and $\{E_m\}$ is the Pauli operator basis $\{I, \sigma_x, \sigma_y, \sigma_z\}$.

Operationally, we reconstruct the $\chi$-matrix by first preparing a complete basis of four input states \cite{shukla2014single} for the victim's qubit: $$\{\rho^{(i)}\} = \{\ket{0}\bra{0}, \ket{1}\bra{1}, \ket{+}\bra{+}, \ket{+i}\bra{+i}\}$$

We then simulate the evolution of each $\rho^{(i)}$ through the attack channel to obtain the corresponding output states, $\rho'^{(i)} = \mathcal{E}(\rho^{(i)})$. The elements of the $\chi$-matrix are then calculated via linear inversion:
\begin{equation}
    \chi_{mn} = \frac{1}{2} \sum_{i} \text{Tr}\left(E_m \rho^{(i)}\right) \text{Tr}\left(E_n^\dagger \rho'^{(i)}\right)
\end{equation}
Resulting $\chi$-matrix serves as a complete and accurate fingerprint of the adversarial process, capturing all its effects, including both coherent rotations and incoherent noise, making it an essential tool for rigorous attack attribution, detection, and defense in multi-tenant quantum platforms.

\subsection{Circuit Approximation Model Justification}

\begin{figure}[h]
\centering
\scalebox{0.9}{
\Qcircuit @C=1.0em @R=0.2em @!R {
\lstick{q_0 : } & \qw & \ctrl{2} & \qw & \qw & \multigate{1}{\mathrm{Measurement}} & \qw & \qw \\
\lstick{q_1 : } & \qw & \qw & \qw & \ctrl{1} & \ghost{\mathrm{Measurement}} & \qw & \qw \\
\lstick{q_2 : } & \gate{R_Y(-\theta)} & \targ & \gate{R_Y(\theta)} & \targ & \qw & \qw & \qw \\
}
}
\caption{Logical circuit model, $\mathcal{E}(\theta)$ seen from Victim's end, representing the effective operation of the crosstalk attack. The $\mathcal{E}(\theta)$, however, is not unitary, rather a general quantum channel described by a CPTP Map.}
\label{fig:model_circuit}
\end{figure}
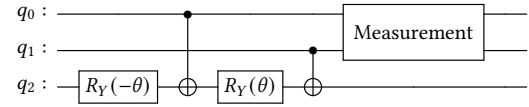

While Quantum Process Tomography (QPT) yields a complete, model-independent description of the quantum channel via the $\chi$-matrix, this characterization remains abstract. To gain an intuitive, operational understanding of the attack, our framework approximates this channel with the effective logical circuit shown in Figure~\ref{fig:model_circuit}. Our choice to model the induced error with a single parameterized rotation, $R_Y(\theta)$ represents a deliberate balance between physical accuracy, interpretability, and computational tractability.\\

\noindent\textbf{Physical Motivation:}  
The choice to model the attack channel as a single-parameter $R_Y(\theta)$ rotation is guided by the nature of physical crosstalk in superconducting qubit platforms. Since our operation space revolves around real valued gates, we chose $R_Y$. \\

\noindent\textbf{Computational Tractability:}  
In principle, an arbitrary single-qubit channel could be modeled as a sequence of three Euler rotations, $R_X(\alpha) R_Y(\beta) R_Z(\gamma)$. However, optimizing over three continuous parameters would dramatically increase the computational burden, particularly given the large parameter sweeps required for the extensive analysis of pulse shapes, amplitudes, and detunings. By reducing the model to a single-parameter fit, our approach makes full-scale process tomography, Kraus reconstruction, and isometry optimization feasible for all simulated cases.
\\
\noindent\textbf{Security and Interpretability:}  
This simplified model offers an additional security benefit: it enables clear and practical detection of adversarial attacks. When the attacker’s qubits are idle (in the $\ket{00}$ state), the sequence of gates on the victim qubit cancels to the identity. Thus, any observed non-zero rotation angle $\theta$ in the fitted model is a direct and interpretable signature of an active crosstalk attack, separating benign drift from adversarial interference. A more expressive model could obscure this key insight, making practical security analysis less transparent.

While the single-parameter $R_Y(\theta)$ model does not capture all possible attack-induced channels, especially if higher-order or multi-axis errors are present, it strikes a crucial balance between physical accuracy, computational efficiency, and interpretability. 

This modeling choice is central to making the full QPT-to-circuit pipeline feasible for security analysis and mitigation in real quantum processors.

It is important to emphasize that, as described in Section~\ref{sec:channels_isometry} (Eq.~\ref{Eq:Kraus}), the effective operation of the attack from the victim’s perspective is always a completely positive, trace-preserving (CPTP) quantum channel. This arises because Adam cannot observe or control the adversarial qubits ($q_0$, $q_1$) and must trace them out after any joint evolution. As a result, even though the global dynamics are unitary, the local action on $q_2$ can be non-unitary, introducing decoherence or mixing. Our modeling with a single-parameter $R_Y(\theta)$ channel captures the dominant coherent effect, but the general CPTP framework is broad enough to encompass not only unitary errors but also more general processes, including decoherence and even measurement-induced disturbance (if, for example, the adversary were to measure their own qubits and discard outcomes). In our present work, we focus on pulse-induced, coherent attacks; however, the QPT and isometry-fitting pipeline would, in principle, detect any such CPTP map, regardless of its physical origin. This highlights the power and generality of the channel-based approach for security analysis and future detection of broader classes of quantum hardware attacks.


\subsection{Isometry Fitting Procedure}

To rigorously determine the parameter $\theta$ that best matches the experimentally reconstructed quantum channel, a simple comparison of Kraus operators is insufficient due to the isometry freedom discussed in subsec.~\ref{sec:channels_isometry}. Our fitting procedure leverages this gauge structure and proceeds as follows:

\begin{enumerate}
    \item \textbf{Derive Theoretical Kraus Operators:}  
    For each value of $\theta$, we analytically derive the four theoretical Kraus operators $\{K_j^{(\text{theory})}(\theta)\}$ for the logical model circuit (Figure~\ref{fig:model_circuit}) by tracing out the adversarial qubits after applying the full unitary. The explicit form is
    \begin{equation} \label{eq:k_all}
    \begin{aligned}
        K_1^{(\text{theory})}(\theta) &= \frac{1}{2}I \\
        K_2^{(\text{theory})}(\theta) &= \frac{1}{2}R_Y(-\theta)\sigma_x R_Y(\theta) \\
        K_3^{(\text{theory})}(\theta) &= \frac{1}{2}\sigma_x \\
        K_4^{(\text{theory})}(\theta) &= \frac{1}{2}R_Y(-\theta)\sigma_x R_Y(\theta)\sigma_x
    \end{aligned}
    \end{equation}

    \item \textbf{Extract Experimental Kraus Operators:}  
    The experimental process matrix ($\chi$) from QPT is converted to the Choi matrix using
    \begin{equation}
        C_{\mathcal{E}} = \sum_{m,n=0}^{3} \chi_{mn} \, E_m \otimes E_n^*
    \end{equation}
    where $\{E_m\}$ is the Pauli basis and $*$ denotes complex conjugation. Diagonalizing $C_{\mathcal{E}}$ yields eigenvalues $\lambda_j$ and eigenvectors $\ket{v_j}$, from which the experimental Kraus operators are constructed as
    \begin{equation}
        K_j^{(\text{exp})} = \sqrt{\lambda_j} \sum_{m=0}^3 \braket{m|v_j} E_m
    \end{equation}
    As discussed in subsec.~\ref{sec:channels_isometry}, this guarantees a valid CPTP channel representation and, in this work, both the model and experiment yield four Kraus operators.
\vspace{1em}
    \item \textbf{Perform Isometry Optimization:}  
    We jointly optimize the logical circuit angle $\theta$ and the unitary isometry $U_{\text{iso}}$ relating the two sets of Kraus operators. The problem is:
    \begin{equation}
        \min_{\theta, U_{\text{iso}}} \sum_{j=1}^{4} \left\| K_j^{(\mathrm{exp})} - \sum_{i=1}^{4} (U_{\text{iso}})_{ji} K_i^{(\mathrm{theory})}(\theta) \right\|_F^2
    \end{equation}
    subject to $U_{\text{iso}}^\dagger U_{\text{iso}} = I$
    where the Frobenius norm $\| \cdot \|_F$ quantifies the operator distance.
    This joint optimization is performed numerically using the Sequential Least Squares Programming (SLSQP) algorithm as implemented in the SciPy library \cite{kraft1988software}. We initialize $U_{\text{iso}}$ as the identity and employ alternating updates: in each iteration, we fix $\theta$ and optimize $U_{\text{iso}}$ (enforcing unitarity via retraction or projection), and then update $\theta$ given the current $U_{\text{iso}}$. This procedure is repeated until convergence is reached, typically requiring only a few cycles due to the low dimensionality of the problem.

    The solution yields the best-fit angle $\theta$, which provides a direct and interpretable quantification of the logical rotation induced by the attack, as well as the optimal isometry $U_{\text{iso}}$, which quantifies the residual gauge mismatch between the model and experiment. The final value of objective function (fit loss) indicates the quality of the model fit: small values confirm that the simple logical model suffices, while larger values signal that true channel includes more complex or incoherent processes beyond a single-axis rotation.
\end{enumerate}
In summary, this procedure yields an interpretable logical error model and a quantitative measure of attack complexity for robust channel diagnosis.

\section{Numerical Characterization of the Adversarial Channel}
\label{sec:numerical_characterization}

This section addresses \hyperlink{rq2}{\textbf{RQ2}} by identifying and characterizing the distinct, asymmetric roles of adversarial qubits during pulse-level crosstalk attacks (Section~\ref{sec:asymmetric_roles}), and completes our investigation of \hyperlink{rq3}{\textbf{RQ3}} by quantifying the robustness and stealthiness of these attacks across various protocols and hardware parameter variations, such as frequency detuning and pulse shapes (Sections~\ref{sec:detuning},~\ref{sec:pulse_shape}).

To achieve this, we performed a systematic numerical analysis. For a range of adversarial parameters, including pulse amplitudes ($A_0, A_1$), frequency detunings, and pulse shapes, we first simulated the full three-qubit system dynamics using a master equation solver. We then performed quantum process tomography (QPT) on the resulting quantum channel acting on the target qubit by tracing out the adversarial qubits. This yielded an experimental chi matrix representation of the induced error channel. To interpret this complex channel in terms of a simple logical operation, we fit its corresponding Kraus operators to a theoretical model of an ideal, single-qubit error channel, as depicted in Figure~\ref{fig:model_circuit}. This fitting procedure minimizes the isometry loss, which quantifies the discrepancy between the simulated attack and the ideal model, and extracts a best-fit rotation angle, $\theta$, representing the effective strength of the logical error. The subsequent analyses explore how this angle and the fit loss depend on the adversary's choices.

\begin{figure*}[!htb]
 \centering
 \includegraphics[width=\linewidth]{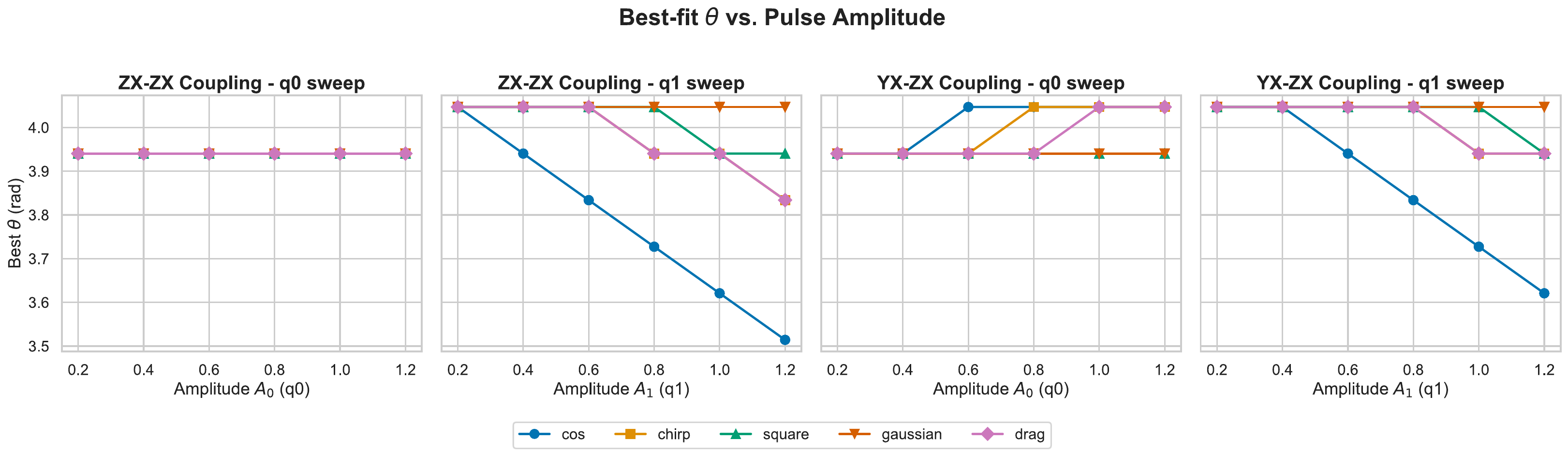}
 \caption{Best-fit rotation angle $\theta$ as a function of the adversarial pulse amplitudes for both ZX-ZX and YX-ZX couplings. The right panels show that the amplitude on $q_1$ (the "driver") directly controls $\theta$. The left panels show that the amplitude on $q_0$ (the "catalyst") has a negligible effect on the rotation angle.}
 \label{fig:theta_vs_amp}
\end{figure*}

\begin{figure*}[!htb]
 \centering
 \includegraphics[width=\linewidth]{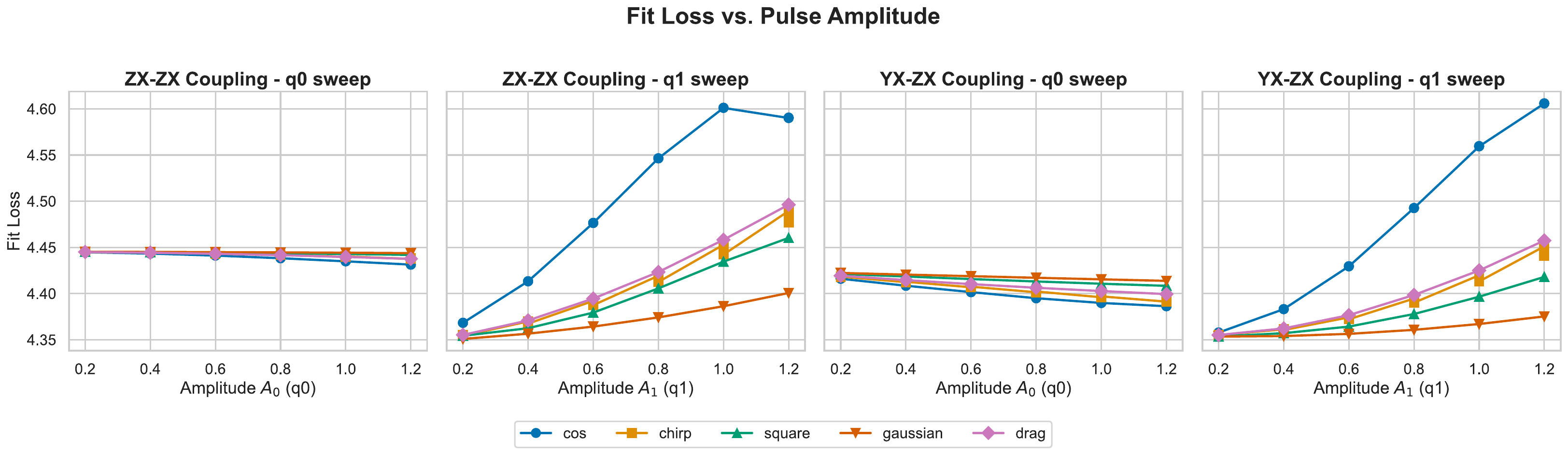}
 \caption{Isometry fit loss as a function of the adversarial pulse amplitudes. The left panels show that increasing the amplitude on $q_0$ (the "catalyst") reduces the fit loss, making the attack more coherent. The right panels show the driver qubit, $q_1$, tends to increase the loss, indicating a more complex, less ideal channel at higher amplitudes.}
 \label{fig:loss_vs_amp}
\end{figure*}

\begin{figure*}[!h]
 \centering
 \includegraphics[width=\linewidth]{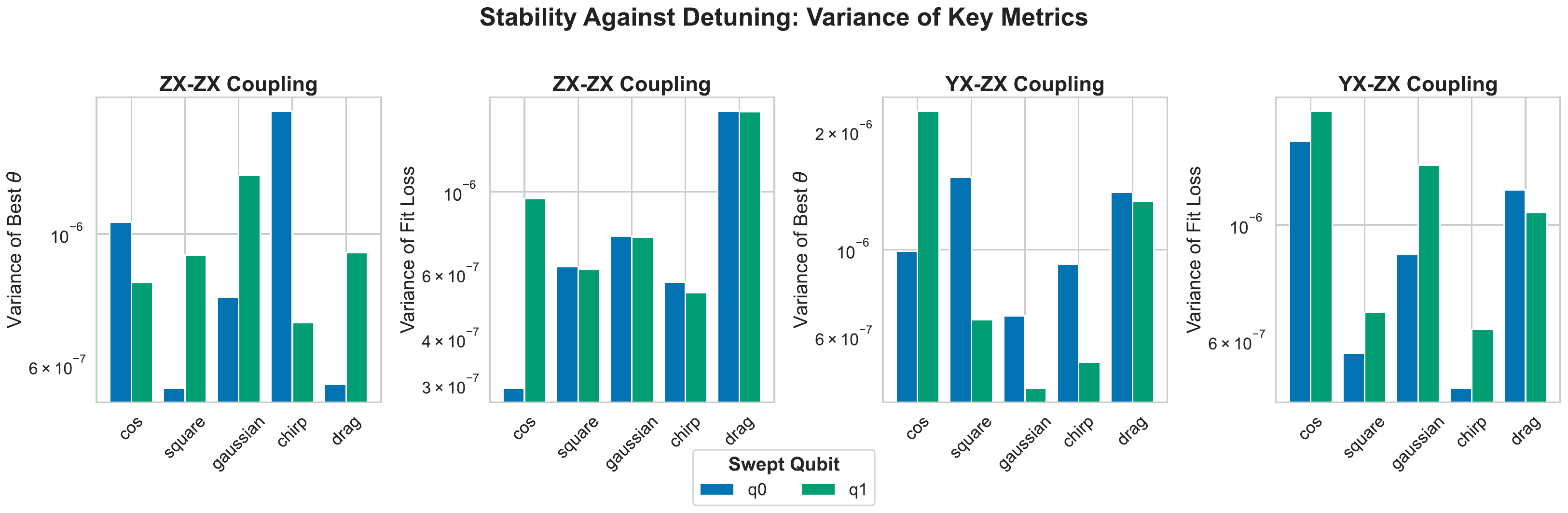}
 \caption{Variance of the best-fit $\theta$ and isometry fit loss across the full detuning range for each pulse shape. The extremely low variance (on the order of $10^{-6}$) for all configurations confirms the attack's high degree of robustness to frequency mismatch.}
 \label{fig:detuning_variance}
\end{figure*}

\subsection{Asymmetric \textit{Driver} vs. \textit{Catalyst} Roles}\label{sec:asymmetric_roles}
Our analysis of the high-influence couplings reveals a critical asymmetric mechanism where the adversarial qubits play fundamentally distinct roles. We identify one qubit as the \textit{driver}, which primarily dictates the magnitude of the attack, and the other as the \textit{catalyst}, which refines and enhances the coherence of the induced error channel.

As shown in Figure~\ref{fig:theta_vs_amp}, by sweeping the pulse amplitude on the \textit{driver} qubit ($q_1$), we observe a direct and significant change in the induced rotation angle, $\theta$. This indicates that pulses applied to the driver qubit strongly control the magnitude of the logical rotation induced on the victim, directly influencing the severity of computational errors. However, as seen in Figure~\ref{fig:loss_vs_amp}, increasing the driver's amplitude also increases the isometry fit loss, suggesting that while the attack becomes stronger, it also introduces more complex dynamics that deviate significantly from our ideal single-axis rotation model in Figure~\ref{fig:model_circuit}.

Conversely, the \textit{catalyst} qubit ($q_0$) exhibits different behavior. Sweeping its amplitude  (left-hand panel of Figure~\ref{fig:theta_vs_amp}) has a negligible effect on the rotation angle $\theta$, but as shown in Figure~\ref{fig:loss_vs_amp}, it significantly reduces the fit loss. This indicates that the catalyst qubit primarily improves the match between the physical attack channel and our theoretical logical circuit model, enhancing coherence without substantially affecting the induced rotation strength. This unique mechanism underscores a sophisticated, two-pronged approach to executing a crosstalk attack.

\subsection{Attack Robustness Against Frequency Detuning}\label{sec:detuning}
A key finding for the practical viability of this attack is its remarkable robustness against frequency errors, a parameter that typically requires precise hardware calibration. To quantify this stability, we analyzed the variance of both the best-fit rotation angle $\theta$ and the isometry fit loss across a wide range of detuning values for all pulse shapes. Our simulations assume closed-system dynamics; incorporating realistic noise and jitter is expected to reduce precision but not eliminate the coherent rotation effect, which remains the dominant error mechanism.

The results are summarized in the bar plot in Figure~\ref{fig:detuning_variance}. The variance for both key metrics is consistently on the order of $10^{-6}$ or smaller, which is negligible. This extremely low variance serves as strong quantitative evidence that the attack's logical effect is insensitive to frequency detuning. From a security perspective, this is a critical result: it implies an adversary can induce a predictable logical error without perfect, real-time knowledge of the system's resonant frequencies. This robustness significantly enhances the attack's stealth, as the induced errors could be easily mistaken for standard calibration drift or random hardware noise.

\subsection{Pulse Shape Sensitivity}\label{sec:pulse_shape}
Further analysis into the impact of different pulse shapes reveals a clear trade-off between attack strength and coherence. We found distinct behaviors among the tested pulse shapes:
\begin{itemize}
    \item \textbf{Cosine and Chirp Pulses:} These shapes induced the most aggressive attack dynamics, resulting in the largest rotation angles. However, they also produced a higher fit loss, indicating a stronger but less coherent, or noisier, attack channel.
    \item \textbf{Square and DRAG Pulses:} These shapes demonstrated more robust and predictable behavior. They induced weaker rotations but had a significantly lower fit loss, corresponding to a more coherent and targeted attack that aligns better with our ideal circuit model.
\end{itemize}
These findings highlight that an adversary can select a pulse shape to balance the trade-off between the magnitude of the induced error and the stealthiness of the attack, offering another layer of control.

\section{Proposed Mitigation Framework}\label{sec:Proposed Mitigation Framework}

While crosstalk attacks can remain stealthy under ordinary use: evading standard gate-level checks and blending into typical hardware noise, our results show that targeted channel tomography or protocol-aware monitoring can reveal a distinctive attack signature, enabling effective detection and mitigation. To counteract the identified attack, we propose a practical, two-tiered mitigation strategy that leverages the distinct signatures of the \emph{attacker-first} versus \emph{victim-first} scenarios. Our approach is based on active monitoring and state reset, rather than attempting to perfectly reverse the adversarial channel.

\subsection*{Tier 1: Anomaly Detection via Canary Circuits}
The significant deviation observed in the attacker-first scenario Figure~\ref{fig:coinflip_default}(a), \ref{fig:coinflip_default}(b) provides a clear, detectable signature. We propose that a quantum system can actively monitor for such attacks by periodically executing simple, fast \textit{canary} circuits, like the Biased Quantum Coin Flip, on idle qubits.

By preparing a known input state (e.g., with $\lambda = \pi/4$) and measuring the output, the system can compare the result to a pre-calibrated, trusted baseline. A large deviation from this baseline, consistent with the attacker-first curve, would serve as a strong indicator of an ongoing integrity attack. The robustness of the XOR protocol to this attack also suggests that the choice of canary circuit is important; a protocol sensitive to small rotational errors is ideal for detection.

\subsection*{Tier 2: Attack Containment via State Reset}
Upon detecting a significant anomaly indicative of an attacker-first attempt, a perfect reversal of the attack is not necessary. A more practical and effective strategy is \textit{attack containment}.

The system's immediate response should be to perform a high-fidelity reset operation on the affected victim qubit. This action effectively nullifies the adversary's attempt to corrupt the initial state. Any subsequent adversarial pulse will now, by definition, fall into the much less damaging \textit{victim-first} regime. Our results show that the impact of a victim-first attack is significantly lower and, in some cases, almost negligible.

This strategy does not eliminate the crosstalk channel itself but contains its threat by forcing it into a less effective operational mode. By combining continuous monitoring with rapid state resets, a multi-tenant quantum system can create a dynamic defense that significantly raises the bar for executing a successful pulse-level crosstalk attack. For more advanced analysis, the detected deviation could be further characterized using our QPT framework to identify a specific \emph{Attack Signature}, informing future hardware design and scheduling policies. In summary, our proposed framework offers a scalable and actionable pathway toward robust crosstalk attack detection and containment in next-generation quantum cloud environments.

\section{Conclusion and Future Work}\label{sec:conclusion}
We have introduced a rigorous, channel-theoretic framework for analyzing and characterizing pulse-level crosstalk attacks in multi-tenant superconducting quantum hardware. By uniting Hamiltonian-level simulation, full quantum process tomography, and a novel isometry-based circuit extraction, we provide the first systematic and broadly applicable pipeline for mapping adversarial pulses to interpretable logical error models, applicable to a wide range of hardware and attack types.
Our results reveal a new asymmetric attack mechanism, where \textit{driver} and \textit{catalyst} qubits play distinct roles in mediating and amplifying logical errors. We further demonstrate that these attacks are both practical and stealthy: they can significantly disrupt diverse quantum protocols, and their effects remain robust even under typical hardware parameter variations.
This work underscores the urgent need to address security at the physical layer of quantum hardware and highlights the value of channel-based analysis for both attack and defense. Our pulse-to-channel-to-circuit methodology enables not only rigorous attack detection but also inspires targeted mitigation via protocol-level anomaly monitoring. Our results motivate the integration of channel-based security analysis into the design, benchmarking, and scheduling policies of future quantum cloud platforms. Future work will focus on deploying this framework for real hardware validation, including platforms such as IQM that provide pulse-level access not currently available on IBM backends, extending it to larger and more complex quantum systems, and refining mitigation strategies into practical, deployable protocols for secure quantum computing at scale.

\bibliographystyle{ACM-Reference-Format}
\bibliography{my_references}


\begin{thebibliography}{22}


\ifx \showCODEN    \undefined \def \showCODEN     #1{\unskip}     \fi
\ifx \showISBNx    \undefined \def \showISBNx     #1{\unskip}     \fi
\ifx \showISBNxiii \undefined \def \showISBNxiii  #1{\unskip}     \fi
\ifx \showISSN     \undefined \def \showISSN      #1{\unskip}     \fi
\ifx \showLCCN     \undefined \def \showLCCN      #1{\unskip}     \fi
\ifx \shownote     \undefined \def \shownote      #1{#1}          \fi
\ifx \showarticletitle \undefined \def \showarticletitle #1{#1}   \fi
\ifx \showURL      \undefined \def \showURL       {\relax}        \fi
\providecommand\bibfield[2]{#2}
\providecommand\bibinfo[2]{#2}
\providecommand\natexlab[1]{#1}
\providecommand\showeprint[2][]{arXiv:#2}

\bibitem[Ash-Saki et~al\mbox{.}(2020)]%
        {ash2020analysis}
\bibfield{author}{\bibinfo{person}{Abdullah Ash-Saki}, \bibinfo{person}{Mahabubul Alam}, {and} \bibinfo{person}{Swaroop Ghosh}.} \bibinfo{year}{2020}\natexlab{}.
\newblock \showarticletitle{Analysis of crosstalk in nisq devices and security implications in multi-programming regime}. In \bibinfo{booktitle}{\emph{Proceedings of the ACM/IEEE International Symposium on Low Power Electronics and Design}}. \bibinfo{pages}{25--30}.
\newblock


\bibitem[Balewski et~al\mbox{.}(2025)]%
        {balewski2025first}
\bibfield{author}{\bibinfo{person}{Jan Balewski}, \bibinfo{person}{Adam Winick}, \bibinfo{person}{Yilun Xu}, \bibinfo{person}{Neel Vora}, \bibinfo{person}{Gang Huang}, \bibinfo{person}{David Santiago}, \bibinfo{person}{Joseph Emerson}, {and} \bibinfo{person}{Irfan Siddiqi}.} \bibinfo{year}{2025}\natexlab{}.
\newblock \showarticletitle{First-principle crosstalk dynamics and Hamiltonian learning via Rabi experiments}.
\newblock \bibinfo{journal}{\emph{arXiv preprint arXiv:2502.05362}} (\bibinfo{year}{2025}).
\newblock


\bibitem[Choudhury et~al\mbox{.}(2024)]%
        {choudhury2024crosstalk}
\bibfield{author}{\bibinfo{person}{Navnil Choudhury}, \bibinfo{person}{Chaithanya~Naik Mude}, \bibinfo{person}{Sanjay Das}, \bibinfo{person}{Preetham~Chandra Tikkireddi}, \bibinfo{person}{Swamit Tannu}, {and} \bibinfo{person}{Kanad Basu}.} \bibinfo{year}{2024}\natexlab{}.
\newblock \showarticletitle{Crosstalk-induced side channel threats in multi-tenant nisq computers}.
\newblock \bibinfo{journal}{\emph{arXiv preprint arXiv:2412.10507}} (\bibinfo{year}{2024}).
\newblock


\bibitem[Giarmatzi et~al\mbox{.}(2023)]%
        {giarmatzi2023multi}
\bibfield{author}{\bibinfo{person}{Christina Giarmatzi}, \bibinfo{person}{Tyler Jones}, \bibinfo{person}{Alexei Gilchrist}, \bibinfo{person}{Prasanna Pakkiam}, \bibinfo{person}{Arkady Fedorov}, {and} \bibinfo{person}{Fabio Costa}.} \bibinfo{year}{2023}\natexlab{}.
\newblock \showarticletitle{Multi-time quantum process tomography on a superconducting qubit}.
\newblock \bibinfo{journal}{\emph{arXiv preprint arXiv:2308.00750}} (\bibinfo{year}{2023}).
\newblock


\bibitem[Grossu(2021)]%
        {grossu2021single}
\bibfield{author}{\bibinfo{person}{IV Grossu}.} \bibinfo{year}{2021}\natexlab{}.
\newblock \showarticletitle{Single qubit neural quantum circuit for solving Exclusive-OR}.
\newblock \bibinfo{journal}{\emph{MethodsX}}  \bibinfo{volume}{8} (\bibinfo{year}{2021}), \bibinfo{pages}{101573}.
\newblock


\bibitem[Ketterer and Wellens(2023)]%
        {ketterer2023characterizing}
\bibfield{author}{\bibinfo{person}{Andreas Ketterer} {and} \bibinfo{person}{Thomas Wellens}.} \bibinfo{year}{2023}\natexlab{}.
\newblock \showarticletitle{Characterizing crosstalk of superconducting transmon processors}.
\newblock \bibinfo{journal}{\emph{Physical Review Applied}} \bibinfo{volume}{20}, \bibinfo{number}{3} (\bibinfo{year}{2023}), \bibinfo{pages}{034065}.
\newblock


\bibitem[Kraft(1988)]%
        {kraft1988software}
\bibfield{author}{\bibinfo{person}{Dieter Kraft}.} \bibinfo{year}{1988}\natexlab{}.
\newblock \showarticletitle{A software package for sequential quadratic programming}.
\newblock \bibinfo{journal}{\emph{Forschungsbericht- Deutsche Forschungs- und Versuchsanstalt fur Luft- und Raumfahrt}} (\bibinfo{year}{1988}).
\newblock


\bibitem[Li et~al\mbox{.}(2022)]%
        {li2022pulse}
\bibfield{author}{\bibinfo{person}{Boxi Li}, \bibinfo{person}{Shahnawaz Ahmed}, \bibinfo{person}{Sidhant Saraogi}, \bibinfo{person}{Neill Lambert}, \bibinfo{person}{Franco Nori}, \bibinfo{person}{Alexander Pitchford}, {and} \bibinfo{person}{Nathan Shammah}.} \bibinfo{year}{2022}\natexlab{}.
\newblock \showarticletitle{Pulse-level noisy quantum circuits with QuTiP}.
\newblock \bibinfo{journal}{\emph{Quantum}}  \bibinfo{volume}{6} (\bibinfo{year}{2022}), \bibinfo{pages}{630}.
\newblock


\bibitem[Maurya et~al\mbox{.}(2024)]%
        {maurya2024understanding}
\bibfield{author}{\bibinfo{person}{Satvik Maurya}, \bibinfo{person}{Chaithanya~Naik Mude}, \bibinfo{person}{Benjamin Lienhard}, {and} \bibinfo{person}{Swamit Tannu}.} \bibinfo{year}{2024}\natexlab{}.
\newblock \showarticletitle{Understanding side-channel vulnerabilities in superconducting qubit readout architectures}. In \bibinfo{booktitle}{\emph{2024 IEEE International Conference on Quantum Computing and Engineering (QCE)}}, Vol.~\bibinfo{volume}{1}. IEEE, \bibinfo{pages}{1177--1183}.
\newblock


\bibitem[Mundada et~al\mbox{.}(2019)]%
        {mundada2019suppression}
\bibfield{author}{\bibinfo{person}{Pranav Mundada}, \bibinfo{person}{Gengyan Zhang}, \bibinfo{person}{Thomas Hazard}, {and} \bibinfo{person}{Andrew Houck}.} \bibinfo{year}{2019}\natexlab{}.
\newblock \showarticletitle{Suppression of qubit crosstalk in a tunable coupling superconducting circuit}.
\newblock \bibinfo{journal}{\emph{Physical Review Applied}} \bibinfo{volume}{12}, \bibinfo{number}{5} (\bibinfo{year}{2019}), \bibinfo{pages}{054023}.
\newblock


\bibitem[Nielsen and Chuang(2010)]%
        {nielsen2010quantum}
\bibfield{author}{\bibinfo{person}{Michael~A Nielsen} {and} \bibinfo{person}{Isaac~L Chuang}.} \bibinfo{year}{2010}\natexlab{}.
\newblock \bibinfo{booktitle}{\emph{Quantum computation and quantum information}}.
\newblock \bibinfo{publisher}{Cambridge university press}.
\newblock


\bibitem[Rudinger et~al\mbox{.}(2021)]%
        {rudinger2021experimental}
\bibfield{author}{\bibinfo{person}{Kenneth Rudinger}, \bibinfo{person}{Craig~W Hogle}, \bibinfo{person}{Ravi~K Naik}, \bibinfo{person}{Akel Hashim}, \bibinfo{person}{Daniel Lobser}, \bibinfo{person}{David~I Santiago}, \bibinfo{person}{Matthew~D Grace}, \bibinfo{person}{Erik Nielsen}, \bibinfo{person}{Timothy Proctor}, \bibinfo{person}{Stefan Seritan}, {et~al\mbox{.}}} \bibinfo{year}{2021}\natexlab{}.
\newblock \showarticletitle{Experimental characterization of crosstalk errors with simultaneous gate set tomography}.
\newblock \bibinfo{journal}{\emph{PRX Quantum}} \bibinfo{volume}{2}, \bibinfo{number}{4} (\bibinfo{year}{2021}), \bibinfo{pages}{040338}.
\newblock


\bibitem[Sarovar et~al\mbox{.}(2020)]%
        {sarovar2020detecting}
\bibfield{author}{\bibinfo{person}{Mohan Sarovar}, \bibinfo{person}{Timothy Proctor}, \bibinfo{person}{Kenneth Rudinger}, \bibinfo{person}{Kevin Young}, \bibinfo{person}{Erik Nielsen}, {and} \bibinfo{person}{Robin Blume-Kohout}.} \bibinfo{year}{2020}\natexlab{}.
\newblock \showarticletitle{Detecting crosstalk errors in quantum information processors}.
\newblock \bibinfo{journal}{\emph{Quantum}}  \bibinfo{volume}{4} (\bibinfo{year}{2020}), \bibinfo{pages}{321}.
\newblock


\bibitem[Shubha and Farheen(2025)]%
        {shubha2025qsec}
\bibfield{author}{\bibinfo{person}{Syed Emad~Uddin Shubha} {and} \bibinfo{person}{Tasnuva Farheen}.} \bibinfo{year}{2025}\natexlab{}.
\newblock \showarticletitle{Pulse-Level Simulation of Crosstalk Attacks on Superconducting Quantum Hardware}. In \bibinfo{booktitle}{\emph{Proceedings of the IEEE Quantum Week (QCE 2025), Workshop on Quantum Computing Security, Privacy and Resilience (Q-SEC)}}. \bibinfo{publisher}{IEEE}.
\newblock
\newblock
\shownote{To appear; preprint: \url{https://doi.org/10.48550/arXiv.2507.16181}}.


\bibitem[Shukla and Mahesh(2014)]%
        {shukla2014single}
\bibfield{author}{\bibinfo{person}{Abhishek Shukla} {and} \bibinfo{person}{TS Mahesh}.} \bibinfo{year}{2014}\natexlab{}.
\newblock \showarticletitle{Single-scan quantum process tomography}.
\newblock \bibinfo{journal}{\emph{Physical Review A}} \bibinfo{volume}{90}, \bibinfo{number}{5} (\bibinfo{year}{2014}), \bibinfo{pages}{052301}.
\newblock


\bibitem[Souza et~al\mbox{.}(2024)]%
        {souza2024regression}
\bibfield{author}{\bibinfo{person}{Leandro~C Souza}, \bibinfo{person}{Bruno~C Guingo}, \bibinfo{person}{Gilson Giraldi}, {and} \bibinfo{person}{Renato Portugal}.} \bibinfo{year}{2024}\natexlab{}.
\newblock \showarticletitle{Regression and Classification with Single-Qubit Quantum Neural Networks}.
\newblock \bibinfo{journal}{\emph{arXiv preprint arXiv:2412.09486}} (\bibinfo{year}{2024}).
\newblock


\bibitem[Tripathi et~al\mbox{.}(2022)]%
        {tripathi2022suppression}
\bibfield{author}{\bibinfo{person}{Vinay Tripathi}, \bibinfo{person}{Huo Chen}, \bibinfo{person}{Mostafa Khezri}, \bibinfo{person}{Ka-Wa Yip}, \bibinfo{person}{EM Levenson-Falk}, {and} \bibinfo{person}{Daniel~A Lidar}.} \bibinfo{year}{2022}\natexlab{}.
\newblock \showarticletitle{Suppression of crosstalk in superconducting qubits using dynamical decoupling}.
\newblock \bibinfo{journal}{\emph{Physical review applied}} \bibinfo{volume}{18}, \bibinfo{number}{2} (\bibinfo{year}{2022}), \bibinfo{pages}{024068}.
\newblock


\bibitem[Wei et~al\mbox{.}(2022)]%
        {wei2022hamiltonian}
\bibfield{author}{\bibinfo{person}{KX Wei}, \bibinfo{person}{Easwar Magesan}, \bibinfo{person}{Isaac Lauer}, \bibinfo{person}{S Srinivasan}, \bibinfo{person}{Daniela~F Bogorin}, \bibinfo{person}{S Carnevale}, \bibinfo{person}{GA Keefe}, \bibinfo{person}{Y Kim}, \bibinfo{person}{D Klaus}, \bibinfo{person}{W Landers}, {et~al\mbox{.}}} \bibinfo{year}{2022}\natexlab{}.
\newblock \showarticletitle{Hamiltonian engineering with multicolor drives for fast entangling gates and quantum crosstalk cancellation}.
\newblock \bibinfo{journal}{\emph{Physical Review Letters}} \bibinfo{volume}{129}, \bibinfo{number}{6} (\bibinfo{year}{2022}), \bibinfo{pages}{060501}.
\newblock


\bibitem[Wood et~al\mbox{.}({[n.\,d.]})]%
        {wood2011tensor}
\bibfield{author}{\bibinfo{person}{Christopher~J Wood}, \bibinfo{person}{Jacob~D Biamonte}, {and} \bibinfo{person}{David~G Cory}.} \bibinfo{year}{[n.\,d.]}\natexlab{}.
\newblock \showarticletitle{Tensor networks and graphical calculus for open quantum systems}.
\newblock \bibinfo{journal}{\emph{arXiv preprint arXiv:1111.6950}} (\bibinfo{year}{[n.\,d.]}).
\newblock


\bibitem[Xu and Szefer(2024)]%
        {xu2024jailbreaking}
\bibfield{author}{\bibinfo{person}{Chuanqi Xu} {and} \bibinfo{person}{Jakub Szefer}.} \bibinfo{year}{2024}\natexlab{}.
\newblock \showarticletitle{Jailbreaking Quantum Computers}.
\newblock \bibinfo{journal}{\emph{arXiv e-prints}} (\bibinfo{year}{2024}), \bibinfo{pages}{arXiv--2406}.
\newblock


\bibitem[Zhao et~al\mbox{.}(2022)]%
        {zhao2022quantum}
\bibfield{author}{\bibinfo{person}{Peng Zhao}, \bibinfo{person}{Kehuan Linghu}, \bibinfo{person}{Zhiyuan Li}, \bibinfo{person}{Peng Xu}, \bibinfo{person}{Ruixia Wang}, \bibinfo{person}{Guangming Xue}, \bibinfo{person}{Yirong Jin}, {and} \bibinfo{person}{Haifeng Yu}.} \bibinfo{year}{2022}\natexlab{}.
\newblock \showarticletitle{Quantum crosstalk analysis for simultaneous gate operations on superconducting qubits}.
\newblock \bibinfo{journal}{\emph{PRX quantum}} \bibinfo{volume}{3}, \bibinfo{number}{2} (\bibinfo{year}{2022}), \bibinfo{pages}{020301}.
\newblock


\bibitem[Zhou et~al\mbox{.}(2025)]%
        {zhou2025characterization}
\bibfield{author}{\bibinfo{person}{Zeyuan Zhou}, \bibinfo{person}{Andrew Ji}, {and} \bibinfo{person}{Yongshan Ding}.} \bibinfo{year}{2025}\natexlab{}.
\newblock \showarticletitle{Characterization and Mitigation of Crosstalk in Quantum Error Correction}.
\newblock \bibinfo{journal}{\emph{arXiv e-prints}} (\bibinfo{year}{2025}).
\newblock


\end{thebibliography}

\end{document}